\documentclass[reprint, amsmath,amssymb,aps, superscriptaddress]{revtex4-1}
\usepackage{dcolumn}
\usepackage{bm}

\usepackage{amssymb}
\usepackage[caption=false]{subfig}
\usepackage{lmodern}
\usepackage{physics}
\usepackage{graphicx}
\usepackage{amsmath}
\usepackage{float}

\begin{document}


\title{Collapse and revival of entanglement between qubits coupled to a spin coherent state}

\author{Iskandar Bahari}
\email{ib640@york.ac.uk}
\author{Timothy P.  Spiller}
 \affiliation{Department of Physics, University of York, York YO10 5DD, United Kingdom }


\author{Shane Dooley}
\author{Anthony Hayes}
\author{Francis McCrossan}
\affiliation{School of Physics and Astronomy, University of Leeds, Leeds LS2 9JT, United Kingdom}%
\date{\today}

\begin{abstract}
 
We extend study of the Jaynes-Cummings model involving a pair of identical two-level atoms (or qubits) interacting with a single mode quantized field. We investigate the effects of replacing the radiation field mode with a composite spin, comprising $N$ qubits, or spin-1/2 particles. This model is relevant for physical implementations in superconducting circuit QED, ion trap and molecular systems. For the case of the composite spin prepared in a spin coherent state, we demonstrate the similarities of this set-up to the qubits-field model in terms of the time evolution, attractor states and in particular the collapse and revival of the entanglement between the two qubits. We extend our analysis by taking into account an effect due to qubit imperfections. We consider a difference (or `mismatch') in the dipole interaction strengths of the two qubits, for both the field mode and composite spin cases. To address decoherence due to this mismatch, we then average over this coupling strength difference with distributions of varying width. We demonstrate in both the field mode and the composite spin scenarios that increasing the width of the `error' distribution increases suppression of the coherent dynamics of the coupled system, including the collapse and revival of the entanglement between the qubits.   


\end{abstract}

\maketitle

\section{\label{sec:level1}Introduction}

Physical applications of an interaction between a two-level system (a qubit) with a field can be observed in many different and interesting quantum systems, such as Rydberg atoms~\cite{haroche1993cavity}, Cooper Pair Boxes~\cite{wallraff2004strong}, Cavity Quantum Electrodynamics ~\cite{berman1994cavity}, trapped ions~\cite{cirac1995quantum} and Circuit QED ~\cite{schuster2007resolving}. A widely used  atom-field interaction model is the Jaynes-Cummings (JC) model. Historically this was introduced by E. Jaynes and F. Cummings~\cite{jcm} and studied to understand the predictions for the state evolution when a two-level atom is coupled to a quantized radiation field, in comparison with the semi-classical interaction model. It was also used to describe the event of spontaneous emission in the quantum theory of radiation. In the present day, because of the wide applicability of this model to a range of qubit-field systems and because of the significant potential for using qubits and fields together in quantum processing, the JC model is used widely as a tool for describing quantum processing ~\cite{you2003quantum, blais2007quantum, an2009quantum} and quantum computing systems ~\cite{azuma2011quantum, mischuck2013qudit, blais2004cavity}.

Following the initial investigations, there have been numerous extended studies of the JC model, conducted with a wide range of physical implementations in mind, to identify the similarities, differences, fundamental features and potential applications of these systems. These studies first led to the discovery of qubit collapse and revival ~\cite{gea1990collapse, gerry2005introductory} and more recently to the discoveries of many new interesting phenomena, such as sudden death of entanglement ~\cite{yu2004finite, qing2008sudden, yu2009sudden}, collapse and revival of qubits entanglement ~\cite{qing2008sudden, jarvis2009dynamics, jarvis2010collapse},  and cat-swapping  ~\cite{jarvis2010collapse}. In this work we study variants of some of these works, particularly those by C. Jarvis \textit{et al.}~\cite{jarvis2009dynamics, jarvis2010collapse} and S. Dooley \textit{et al}.~\cite{dooley2013collapse}. We propose a new model that incorporates some selected features of the earlier works into a different hybrid system with qubits, that should also be a candidate for quantum information and processing applications. This model demonstrates the appearance of a range of phenomena with potential application, including the collapse and revival of Rabi oscillations, the attractor and the cat states of the system, as well as the dynamics of the entanglement between the two initial qubits in the system.

\section{Two Qubit Jaynes-Cummings System}\label{NQJCM}

A multi-qubit Jaynes-Cummings Hamiltonian~\cite{tavis1968exact} for the interaction of a bosonic field mode with $M_{q}$ qubits is written in the form of
\begin{equation}
\widehat{H}=\hbar\omega(\hat{a}^{\dagger}\hat{a})+\frac{\hbar}{2}\sum_{i=1}^{M_q}{\Omega_i\hat{\sigma}^{z}_i} +\hbar\sum_{i=1}^{M_q}{\lambda_i (\hat{a}^{\dagger}\hat{\sigma}^{-}_i+\hat{a}\hat{\sigma}^{+}_i)}
\end{equation}

\noindent where $\hat{\sigma}_i^{z}=\left|e_i\right\rangle \left\langle e_i\right|-\left|g_i\right\rangle \left\langle g_i\right|$, $\hat{\sigma}_i^{+}=\left|e_i\right\rangle \left\langle g_i\right|$ and $\hat{\sigma}_i^{-}=\left|g_i\right\rangle \left\langle e_i\right|$ are the qubit operators, $\hat{a}^{\dagger}$ and $\hat{a}$ are respectively the creation and annihilation operators for a photon with frequency $\omega$, and $\lambda_i$ is the cavity-qubit$_i$ coupling constant. Each qubit labeled by $i$ has ground $\left|g_i\right\rangle$ and excited $\left|e_i\right\rangle$ states with energy $\epsilon_{g,i}$ and $\epsilon_{e,i}$ respectively, where $\hbar\Omega_i=\epsilon_{e,i}-\epsilon_{g,i}$. Initially we consider the case of identical qubits, and with the atomic and the field inversion frequencies at resonance and the field having a uniform coupling with each qubit, so that for all $i$, $\omega=\Omega_i$ and $\lambda=\lambda_i$. 

Considering the case of two qubits ($M_{q}=2$) interacting with a coherent state of the field, the initial system has the form of $\left|\Psi_2 (0)\right\rangle=\left|\psi_2\right\rangle\left|\alpha\right\rangle$, where the qubits are in state $\left|\psi_2\right\rangle=C_{ee} \left|ee\right\rangle+C_{eg} \left|eg\right\rangle+C_{ge} \left|ge\right\rangle+C_{gg} \left|gg\right\rangle$ and the field is in state $\left|\alpha\right\rangle=e^{{-\lvert\alpha\rvert}^2/2}\sum_{n=0}^{\infty}\frac{\alpha^{n}}{\sqrt{n!}}\left|n\right\rangle$. Here $|gg\rangle$ is shorthand for $\left|g_1\right\rangle \left|g_2\right\rangle$ etc., the $|n\rangle$ are the number (energy) eigenstates of the field, $\alpha=\sqrt{\bar{n}}e^{-i\theta}$ and $\bar{n}$ is the average photon number in the coherent state. The time evolution of this system $\left|\Psi_2 (t)\right\rangle$ has been obtained and discussed previously ~\cite{chumakov1995collective}, where the system undergoes collapse and revival of Rabi oscillations and in particular, the collapse in the Rabi oscillations is observed at $t_c\approx\frac{2}{\lambda}$ and then the revival occurs at a later time $t_r\approx\frac{2 \pi\sqrt{\bar{n}}}{\lambda}$. 

\begin{figure}[!htb]
\centering
\includegraphics[scale=.40]{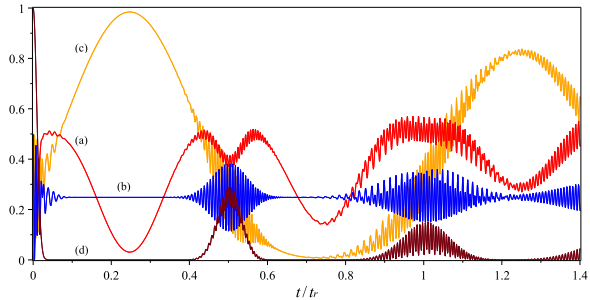}
\caption{\label{fig:aps1} Time evolution for two-qubit Jaynes-Cummings system with the initial two-qubit state $\frac{1}{\sqrt{2}}(\left|ee\right\rangle+\left|gg\right\rangle)$, $\bar{n}=36$ and the initial phase of the radiation field $\theta=0$. (a) the linear entropy of the qubits. (b) the probability of the two-qubit state $\left|ee\right\rangle$. (c) the probability of being in the two qubit attractor state $\left|\psi_{2,att}^+\right\rangle$. (d) tangle to quantify entanglement between the qubits. } 
\end{figure}

In a detailed analysis of the one-qubit case, J. Gea-Banacloche  ~\cite{gea1991atom} showed that, half way to the first revival, every initial qubit state approaches the single-qubit attractor 
$\left|{\psi^{+}_{1,att}}\right\rangle $ and between the first and second revivals it approaches $\left|{\psi^{-}_{1,att}}\right\rangle $, with
\begin{equation} \label{eqatt1qb}
\left|{\psi^\pm_{1,att}}\right\rangle = \frac{1}{\sqrt{2}} \left(e^{-i\theta}\left|e\right\rangle \pm i\left|g\right\rangle \right) \; .
\end{equation}
C.Jarvis $et.al$ extended this analysis ~\cite{jarvis2010collapse,jarvis2009dynamics} and have shown that for the case of the two-qubit JC model, in the large $\bar{n}$ approximation, there are three components to the full time-evolving state. Provided that one component vanishes, the other two combine at the appropriate times (half way to revival) to a product state of the field with two-qubit  states that are the unentangled attractor states ~\cite{shore1993jaynes} of the system. These are products of the single-qubit attractor states   and take the form
\begin{equation} \label{eqatt}
\left|{\psi^\pm_{2,att}}\right\rangle = \frac{1}{2} \left(e^{-2i\theta}\left|ee\right\rangle \pm ie^{-i\theta}(\left|eg\right\rangle + \left|ge\right\rangle) - \left|gg\right\rangle\right) \; .
\end{equation}
Their occurrence thus depends on the initial state of the two qubits. This is in contrast to the one-qubit case in which all initial qubit states lead to a Schr\"{o}dinger cat state at the attractor times ~\cite{gea1991atom}. The vanishing of one part of the initial state therefore quantifies the `basin of attraction' $i.e.$ all the initial states that lead to the attractor states. These states lie in the symmetric subspace and are defined as
\begin{equation}\label{eqbasin}
\left|\psi_{2,basin}\right\rangle=a(e^{-i\theta} \left|ee\right\rangle+ e^{i\theta}\left|gg\right\rangle)+ \sqrt{\frac{1}{2}-\lvert a \rvert^2} (\left|eg\right\rangle+\left|ge\right\rangle)
\end{equation}
\noindent where $\theta$ is the initial phase of the radiation field and $a$ satisfies $0\leq\lvert a \rvert \leq\frac{1}{\sqrt{2}}$. Jarvis $et.al$ have also generalized the `basin of attraction' for  all the fully symmetrized $M_q$ qubit attractor states ~\cite{jarvis2009dynamics}. 

Fig.~\ref{fig:aps1} illustrates a few interesting quantities of such interactions for an initial state with $C_{ee}=C_{gg}=\frac{1}{\sqrt{2}}$ and $C_{eg}=C_{ge}=0$. The linear entropy $S^L_q=1-\Tr(\rho^2_q(t))$ associated with the reduced density matrix of the two qubits, when the field has been traced out, $\rho_q(t)=\Tr_f(\ket{\Psi_2(t)}\bra{\Psi_2(t)})$ is plotted as line (a) as a function of time. Rabi oscillations of the time evolution $\left|\Psi_2 (t)\right\rangle$ are plotted as line (b)  by calculating $\sum_{n=0}^{\infty}\lvert\bra{ee,n}\ket{\Psi_2 (t)}\rvert^2$ where  $\bra{ee,n}$ corresponds to the state with both qubits in their excited state and with $n$ photons in the field mode. The probability of the two qubits being in the attractor state  of equation (\ref{eqatt}) is also calculated by $\left\langle {\psi^+_{2,att}}|\rho_q(t)|{\psi^+_{2,att}}\right\rangle$ and shown as line (c).  Another interesting phenomenon  observed from such interactions is the collapse and revival of entanglement between the two qubits ~\cite{jarvis2009dynamics}. As opposed to the one-qubit case, entanglement can also be measured between the two qubits, not just between the qubits and the field. The dynamics of such entanglement is defined by the mixed state tangle, $\tau(\rho_q)$ ~\cite{wootters1998entanglement,wootters2001entanglement} and is depicted as line (d) in Fig.~\ref{fig:aps1} . This is calculated from the (generally mixed) two-qubit reduced density matrix, which is calculated from the full system state by tracing out the field.

\begin{figure}[!htb]
\centering
\includegraphics[scale=.40]{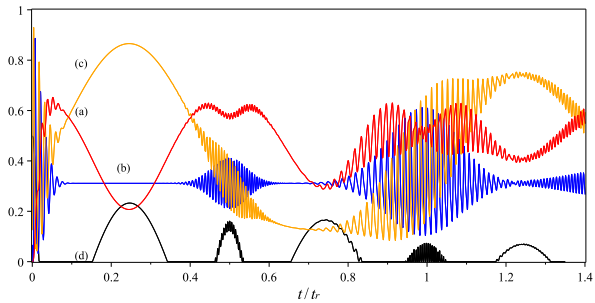}
\caption{\label{fig:aps2}Time evolution for two qubits Jaynes-Cummings system with the initial qubits state $\frac{1}{\sqrt{15}}\left|ee\right\rangle+\sqrt{\frac{14}{15}}\left|gg\right\rangle$, $\bar{n}=36$ and the initial phase of the radiation field $\theta=0$. (a) the linear entropy of the qubits. (b) the probability of the two qubit state $\left|ee\right\rangle$. (c) the probability of being in the two qubit attractor state  $\left|\psi_{2,att}^+\right\rangle$. (d) concurrence to quantify entanglement between the qubits. }
\end{figure}

Fig.~\ref{fig:aps2} shows similar parameters as in Fig.~\ref{fig:aps1} but for the case of an initial state  lying outside the `basin of attraction' described by equation (\ref{eqbasin}). In this case, instead of the phenomenon of collapse and revival, we observe sudden birth or death in the entanglement between the two qubits ~\cite{yu2004finite, qing2008sudden, yu2009sudden}. It is represented by a measurement called `concurrence', $C(\rho)=\sqrt{\tau(\rho_q)}$ and plotted as line (d).

\section{Two Qubits-Composite Spin System}\label{2QBS}

There is significant interest and physical motivation to consider models where the field mode is replaced by a composite of $N$ spins. With a physical implementation such as Circuit QED \cite{mcdermott2005simultaneous, niskanen2007quantum, neeley2010generation, neeley2009emulation, dicarlo2010preparation, reed2012realization,kirchmair2013observation}, one could consider a system of multiple superconducting qubits all equally coupled to one or more microwave field modes operated in their qubit limit (zero or one excitation). Similarly in ion trap or atomic systems \cite{bohnet2016quantum, reiter2017autonomous, schindler2013quantum, gorman2017quantum} one could consider a system of ions or atoms coupled to various cavity modes operated in their qubit limit. With increasing control over the nature of molecular fabrication \cite{jones2009magnetic,simmons2010magnetic}, one can envisage designer molecules where one or more specific spins couple to a system of $N$ other spins. Another possibility is realisation of  nuclear spin control through an interaction of an electron spin with nuclear spin for quantum computation using semiconductor quantum dots \cite{reilly2010exchange, chekhovich2013nuclear}.

The interactions between a single (specified) qubit with a collection of $N$ spin-$\frac{1}{2}$ particles or qubits (collectively called a `big spin', or `Dicke spin', or composite spin) have been studied by S. Dooley and collaborators ~\cite{dooley2013collapse}. They have shown the similarities of this system with a standard JC model, leading to a method for producing   `spin cat states' of the composite spin. Analogous to use of an initial coherent state in the field mode case, an initial spin coherent state is used in the composite spin case. The correspondence between the two models holds in the limit of $N\rightarrow\infty$ where a scaled spin coherent state $\left|\frac{\zeta}{\sqrt{N}}\right\rangle_N$ is equivalent to an oscillator coherent state of amplitude $\zeta$, where $\zeta=\left|\zeta\right|e^{-i\phi}$ is a complex number. The subscript $N$ is used to differentiate a system with a composite spin from a state of the field mode. The spin coherent state can be written in terms of the well known Dicke states $\left|\frac{N}{2},n-\frac{N}{2}\right\rangle_N$ as:

\begin{equation} \label{dickiebs} {\textstyle
\left|\frac{\zeta}{\sqrt{N}}\right\rangle_N=\sum\limits^{N}_{n=0} C_n \left|\frac{N}{2}, n-\frac{N}{2}\right\rangle}
\end{equation}

\noindent where 
\begin {equation}
C_n=\frac{1}{\left(1+\frac{|\zeta|^{2}}{N}\right)^{N/2}}\sqrt{\frac{N!}{(N-n)!n!}}\left(\frac{\zeta}{\sqrt{N}}\right)^n. 
\end {equation}

\noindent Alternatively, the spin coherent state can be written as a product of $N$ identical single spin pure states. In this form it is clear that preparation of such an initial state of the composite spin in any relevant physical implementation should be relatively straightforward, because each spin in the composite can be prepared separately and simultaneously---there is no initial entanglement required.

The Hamiltonian for  a system of $M_{q}$ multiple (specified) qubits interacting with a composite spin is given by:
\begin{equation} \label{hamilbs}
\widehat{H}=\hbar\omega(\hat{J}^{z}+\frac{N}{2})+\frac{\hbar}{2}\sum\limits_{i=1}^{N_q}{\Omega_i\hat{\sigma}^{z}_i} +\frac{\hbar}{\sqrt{N}}\sum_{i=1}^{N_q}{\lambda_i (\hat{J}^{+}\hat{\sigma}^{-}_i+\hat{J}^{-}\hat{\sigma}^{+}_i)}
\end{equation}
\noindent where $\hat{J}^{z}\equiv\frac{1}{2}\sum\limits^N_{k=0}\hat{\sigma}^{z}_k$ and $\hat{J}^{\pm}\equiv\frac{1}{2}\sum\limits^N_{k=0}\hat{\sigma}^{\pm}_k$ are the `big spin' operators and $\hat{\sigma}_k^{z}=\left|e_k\right\rangle \left\langle e_k\right|-\left|g_k\right\rangle \left\langle g_k\right|$ are the individual qubit  operators. The constant term $\frac{\omega N}{2}$ is introduced so that the ground state eigenvalue of the composite spin Hamiltonian $\hat{J}^{z}$ is zero. 

We consider a qubits-composite spin system evolving by Hamiltonian (\ref{hamilbs}) with $M_q=2$, and with the composite spin initially in a spin coherent state described by equation (\ref{dickiebs}). With identical initial qubits, the atomic and composite spin inversion frequencies are taken to be at resonance so that $\omega=\Omega_1=\Omega_2$, and the dipole-interaction strengths between these are   uniform with $\lambda=\lambda_1=\lambda_2$. Given a maximally entangled initial state of the two specified qubits
\begin{equation}
\left|\psi_2\right\rangle=\frac{1}{\sqrt{2}} \left|ee\right\rangle+\frac{1}{\sqrt{2}}\left|gg\right\rangle
\end{equation}  

\noindent the initial system state has the form of 

\begin{equation}
\left|\Psi_2 (0)\right\rangle_N=\left|\psi_2\right\rangle\left|\frac{\zeta}{\sqrt{N}}\right\rangle_N 
\end{equation}

\noindent and the revival time of the system is approximated as
\begin{align}
t_{r}=\frac{2\pi\sqrt\frac{{N|\zeta|^2}}{N+|\zeta|^2}}{\lambda \big(1-\frac{3|\zeta|^2}{2(N+|\zeta|^2)}-\frac{N+|\zeta|^2}{N|\zeta|^2}+\frac{|\zeta|^2}{4N^2}\frac{N(|\zeta|^2-1)+|\zeta|^2(N-1)}{{(N+|\zeta|^2)}^2}\big)}
\end{align}

\noindent which is consistent with the findings of the study on the one qubit-composite spin interaction system ~\cite{dooley2014quantum}.

To exhibit the similarity with the analogous phenomena in the field mode case (Fig. 1) we plot the time evolutions of this interacting system in Fig.~\ref{fig:BS00all}.  Line (a) is for the linear entropy $S_{q}^L(t)$ of the two qubits, calculated from the two-qubit  reduced density matrix tracing the full state over the composite spin and line (b) to show the probability of the two qubits being in state $\lvert ee\rangle$. 

One notable feature of this time evolution is that at time $\frac{1}{4}t_r$ the probability of being in the attractor state approaches a maximum and the entropy approaches zero indicating no entanglement between the the two qubits and the composite spin. Clearly at this particular time, the system factorises into a two-qubit  part and a composite spin part ~\cite{gea1990collapse}. Similar to the two qubits JC model, $\frac{1}{4}t_r$ is the attractor time and the state of the qubits at this time can be written in as equation ~(\ref{attbs}) with the composite spin initial phase, $\phi$. The probability of being in this state is plotted as line (c) in Fig.~\ref{fig:BS00all}.  
\begin{equation}\label{attbs}
\left|{\psi^\pm_{2,att}}\right\rangle_N = \frac{1}{2} \left(e^{-2i\phi}\left|ee\right\rangle \pm ie^{-i\phi}(\left|eg\right\rangle + \left|ge\right\rangle) - \left|gg\right\rangle\right).
\end{equation}

\begin{figure}[]
\centering
\includegraphics[scale=.50]{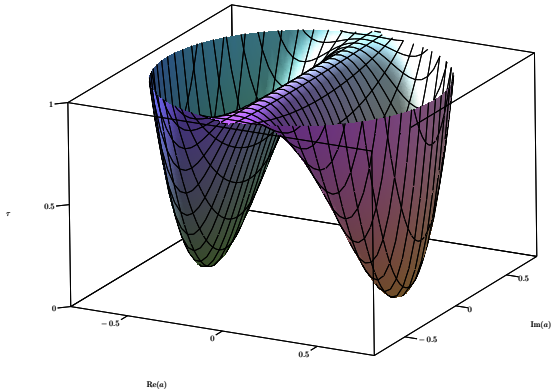}
\caption{\label{fig:basin3D} Tangle value for the states that lie within the `basin of attraction' for different values of $a$. } 
\end{figure}

\begin{figure}[H]
\centering
\includegraphics[scale=.40]{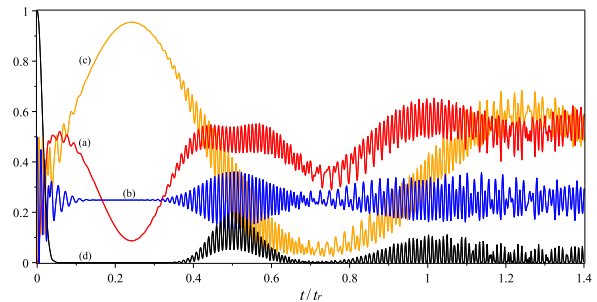}
\caption{\label{fig:BS00all} Time evolution for two qubits-composite spin system with the initial qubits state $\frac{1}{\sqrt{2}}(\left|ee\right\rangle+\left|gg\right\rangle)$, $|\zeta|^2=25$, $N=150$ and the big spin's initial phase, $\phi=0$. (a) the linear entropy of the qubits. (b) the probability of the two-qubit state $\left|ee\right\rangle$. (c) the probability of being in the two qubit attractor state  $\left|\psi_{2,att}^+\right\rangle$. (d) tangle to quantify entanglement between the qubits. } 
\end{figure}

\noindent These states are also dependent on the initial state of the qubits and in the limit of $N$ approaching infinity, we recover similar `basin of attraction' as equation (\ref{eqbasin}) for this interaction system. Fig.~\ref{fig:basin3D} exhibits the value of tangle for the states in the `basin of attraction' and the plot shows that tangle is zero ($\tau=0$) at just two points, indicating only two product states in the `basin of attraction'.

A second dip in the entropy can be seen at time $\frac{3}{4}t_r$ of Fig.~\ref{fig:BS00all} but here the probability of being in the attractor state $\left|{\psi^+_{2,att}}\right\rangle$ goes to zero. This indicates that the two qubits have once again (partially) disentangled themselves from the composite spin. At this point, the qubits are again in an approximately pure state, but in the orthogonal attractor state $\left|{\psi^-_{2,att}}\right\rangle_N$. 

Another interesting feature of this system is the dynamics in the entanglement between the two qubits. We observe the time evolution of the system's mixed state tangle calculated from $\rho_q (t)$ and we plot line (d) to represent the tangle $\tau(t)$ that quantifies entanglement between the qubits at time $t$. Evidently, the two-qubit entanglement collapses and revives and this is similar to the phenomena discovered by C. Jarvis \textit{et al.} ~\cite{jarvis2008collapsebus} and Rodrigues \textit{et al.} ~\cite{rodrigues2004arrays}. It can be seen that at the attractor times of $\frac{1}{4}t_r$ and $\frac{3}{4}t_r$, the tangle values are zero, illustrating no entanglement exists between the qubits. Clearly at these times, the system has gone from having entanglement to having no entanglement, neither between the two qubits and the composite spin, nor between the two qubits themselves. Also, since the state of the full system is pure at all times, the composite spin must also be in a pure state such that the quantum information in the initial (entangled) state of the two qubits has been swapped into the state of the composite spin, which is in a spin cat state. 

If a perfect revival of the whole system state were to occur at the revival time, it would be expected that the entanglement between the two qubits would 
 be exactly the same as it was at the beginning of the interaction, so $\tau(0)=\tau(\frac{t_r}{2})$. During the evolution to the attractor time the quantum information in the state of the two qubits is swapped with that of the composite spin. If there was a perfect revival this process would reverse so that  $\tau$ returns to its initial value at $\frac{t_r}{2}$. However, a trade-off in tangle for entropy prevents a perfect revival from happening. The tangle and entropy for a mixed two-qubit system are approximately inversely proportional to each other ~\cite{munro2001maximizing}, therefore maximum entanglement can only occur for a completely pure state. At the revival time the two-qubit entropy does not return to zero, so the initial entanglement cannot completely revive, although as shown in Fig.~\ref{fig:BS00all} there is a marked revival in the tangle. A further point to note is that the value of the tangle in between revivals collapses and remains near zero until the revival, indicating an occurrence of a phenomenon analogous to the `collapse and revival of entanglement'~\cite{jarvis2008collapsebus} seen in the case of two qubits coupled to a field mode.

We plot Fig. 5 to show the time evolutions of a two qubits-composite spin system for the case of initial states from outside the `basin of attraction'. It can be seen that there are concurrence peaks observed at $\frac{1}{2}t_r$ and $t_r$ indicating the existence of entanglement between the qubits. However, these happen with a lower maximum and display a sudden death and birth behaviour, rather than collapse and revival. This is analogous to the `sudden death of entanglement' that has been seen in the case of two qubits coupled to a field mode ~\cite{yu2004finite,yu2009sudden, jarvis2009dynamics}. Concurrence peaks can also be observed at $\frac{1}{4}t_r$ and $\frac{3}{4}t_r$, which is in contrast to the case of the initial state  inside the `basin of attraction', where no entanglement occurs at  these particular times. Also, for a perfect revival there would be a similar expectation of $\tau(0)=\tau(\frac{t_r}{2})$. However,  as previously the revival of entanglement is not complete as the entropy of the two qubits does not return to zero at the revival time. These results are quantitatively similar to the case of the two qubits-field mode system.

\begin{figure}[!htb]
\centering
\includegraphics[scale=.40]{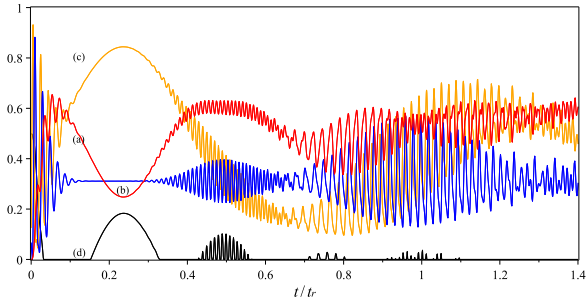}
\caption{\label{fig:BS00allnonmax} Time evolution for two qubits-composite spin system with the initial qubits state $\frac{1}{\sqrt{15}}\left|ee\right\rangle+\sqrt{\frac{14}{15}}\left|gg\right\rangle$, $|\zeta|^2=25$, $N=150$ and a composite spin initial phase of $\phi=0$. (a) the entropy of the qubits. (b) the probability of the two qubit state $\left|ee\right\rangle$. (c) the probability of being in the two qubit attractor state  $\left|\psi_{2,att}^+\right\rangle$. (d) concurrence to quantify entanglement between the two qubits. } 
\end{figure}

\section {Two Qubits-Composite Spin System with Decoherence Effects}

Decoherence is the transformation of a quantum-mechanical superposition state into a classical statistical mixture over time, as a result of the quantum system interacting with an external environment or noise~\cite{gerry2005introductory}. All measurements discussed in the previous sections are observed without considering any `error' or decoherence effects on the system. We are interested to examine an interacting system with decoherence effects, by considering an error that resulted from some mismatch in dipole-interaction strength between the qubits and the composite spin. This means that we are observing a system of two qubits with some `errors' in  the value of $\lambda$. We start by letting $\lambda_1=\lambda_2+\delta$ in Hamiltonian ~(\ref{hamilbs}) (for the case of $M_q=2$), where $\delta$ is the difference between the two dipole-interaction strengths. In order to study a realistic model for this mismatch as a source of decoherence, that obeys the `central limit theorem', we evaluate the density matrix of the qubits for a distribution of the $\delta$ values. We consider our model with errors described by a Gaussian distribution centred at $\delta=0$ and with a width of $\Delta$. The error distribution is then given by

\begin{align}
f\left(\delta |0, \Delta\right)=\frac{1}{\Delta\sqrt{2\pi}}e^{-\frac{\delta^2}{2\Delta^2}} \; .
\end{align}
\\Our aim is to evaluate the density matrix of the qubits given a distribution of $\delta$ values. Analytically, we can do so by averaging the density matrix over the errors such that

\begin{align}
\rho_q (t)=\int^{\infty}_{-\infty}d\delta f\left(\delta |0, \Delta\right) \rho_q(t,\delta) \; .
\end{align}

\begin{figure}[h]
	\subfloat[$\Delta=0.1$]{%
		\includegraphics[width=0.85\columnwidth]{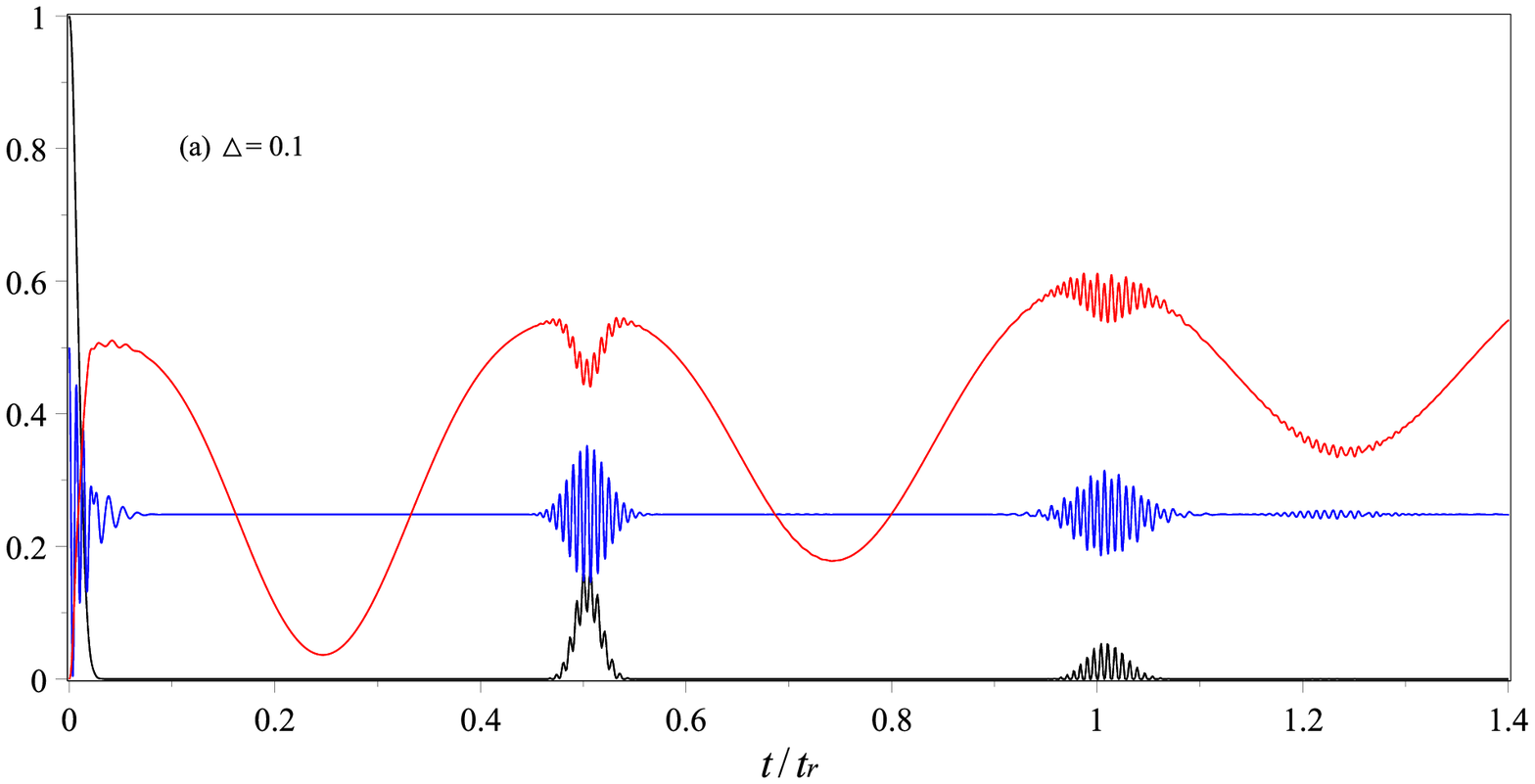} %
		}
		
	\subfloat[$\Delta=0.3$]{%
		\includegraphics[width=0.85\columnwidth]{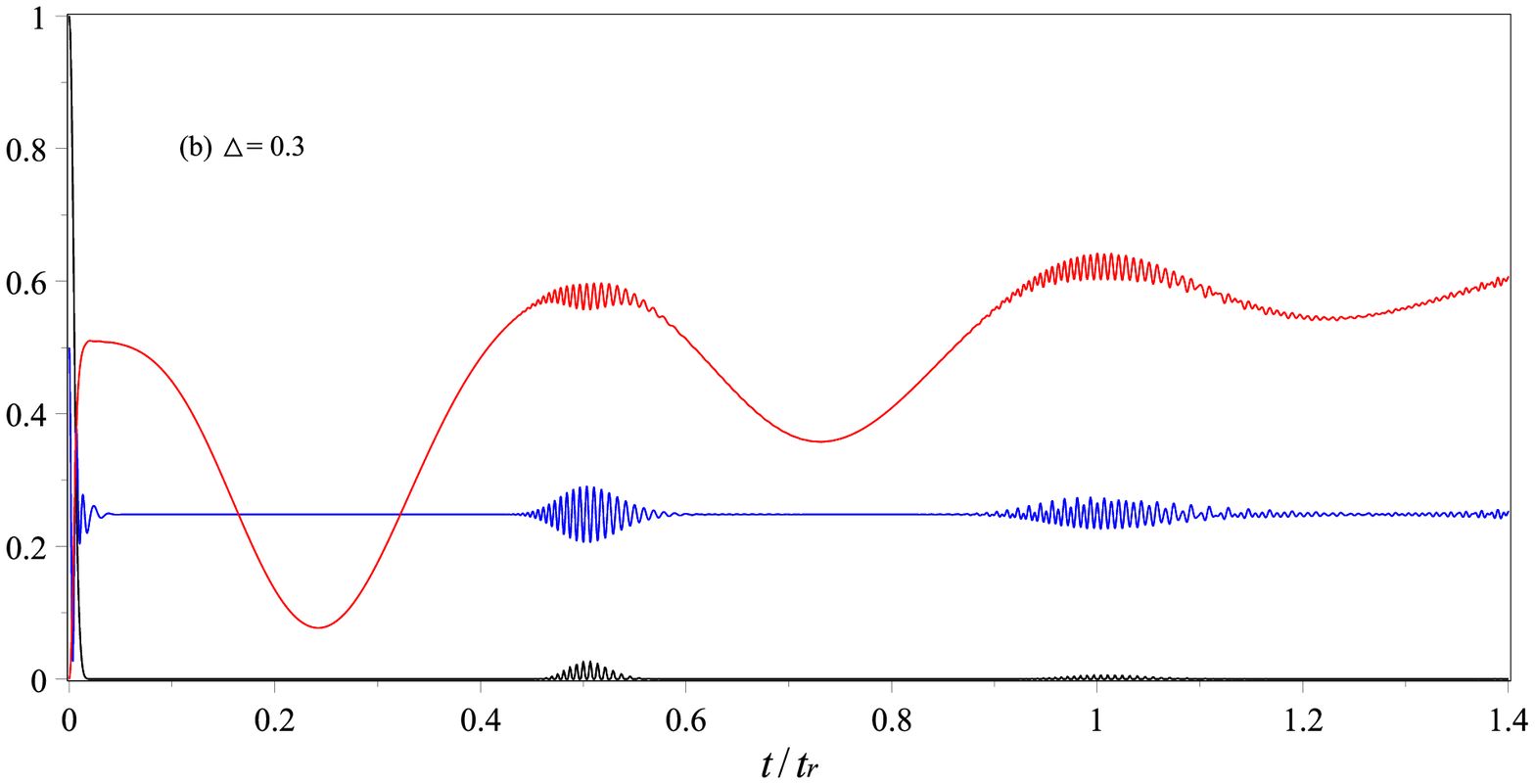}%
		}
		
	\subfloat[$\Delta=0.5$]{%
		\includegraphics[width=0.85\columnwidth]{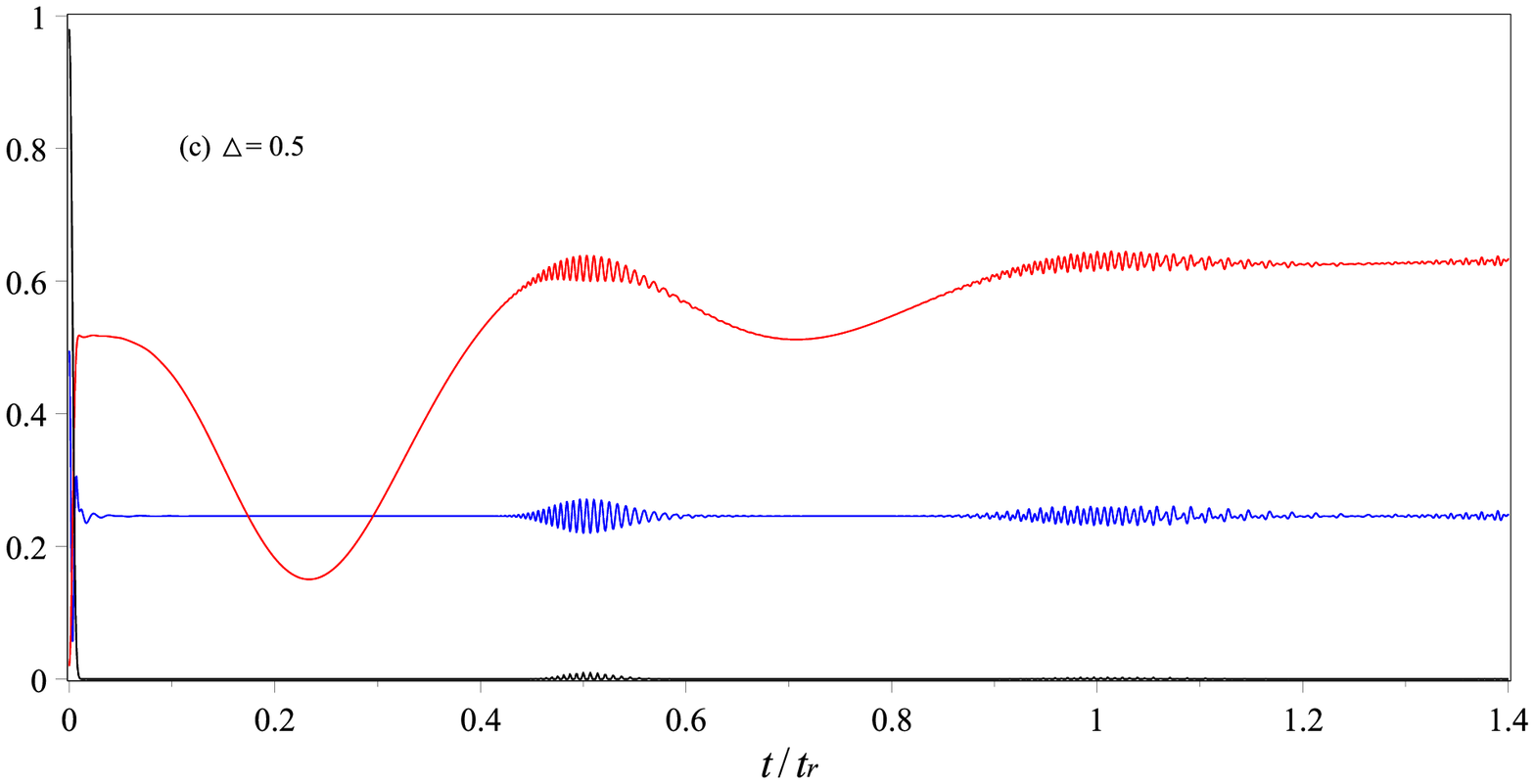}%
		}
	
\caption{Plots comparing qubits linear entropies (red), probability of the two qubits state $\left|ee\right\rangle$ (blue) and tangle (black) for two qubits-field mode models of initial qubit state $\frac{1}{\sqrt{2}}(\left|ee\right\rangle+\left|gg\right\rangle)$, $\bar{n}=36$ and the initial phase of the radiation field $\theta=0$, with decoherence effects.  Figure (a) shows the differences in the system with $\Delta = 0.1$, (b) shows the differences in the system with $\Delta = 0.3$ and (c) shows the
differences in the system with $\Delta = 0.5$.}
\label{fig:DCGCS}
\end{figure}

\begin{figure}[h]
	\subfloat[$\Delta=0.1$]{%
		\includegraphics[width=0.85\columnwidth]{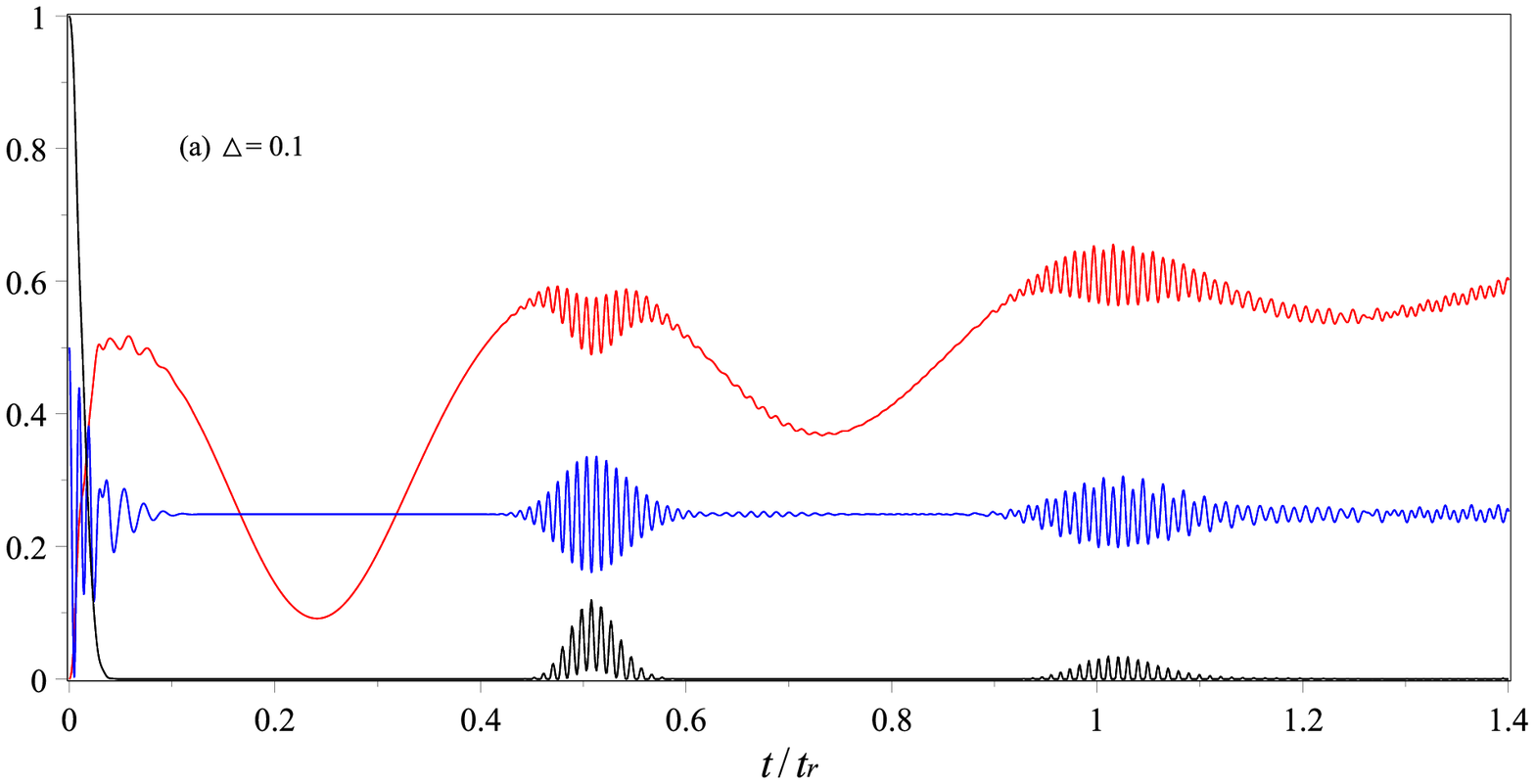} %
		}
		
	\subfloat[$\Delta=0.3$]{%
		\includegraphics[width=0.85\columnwidth]{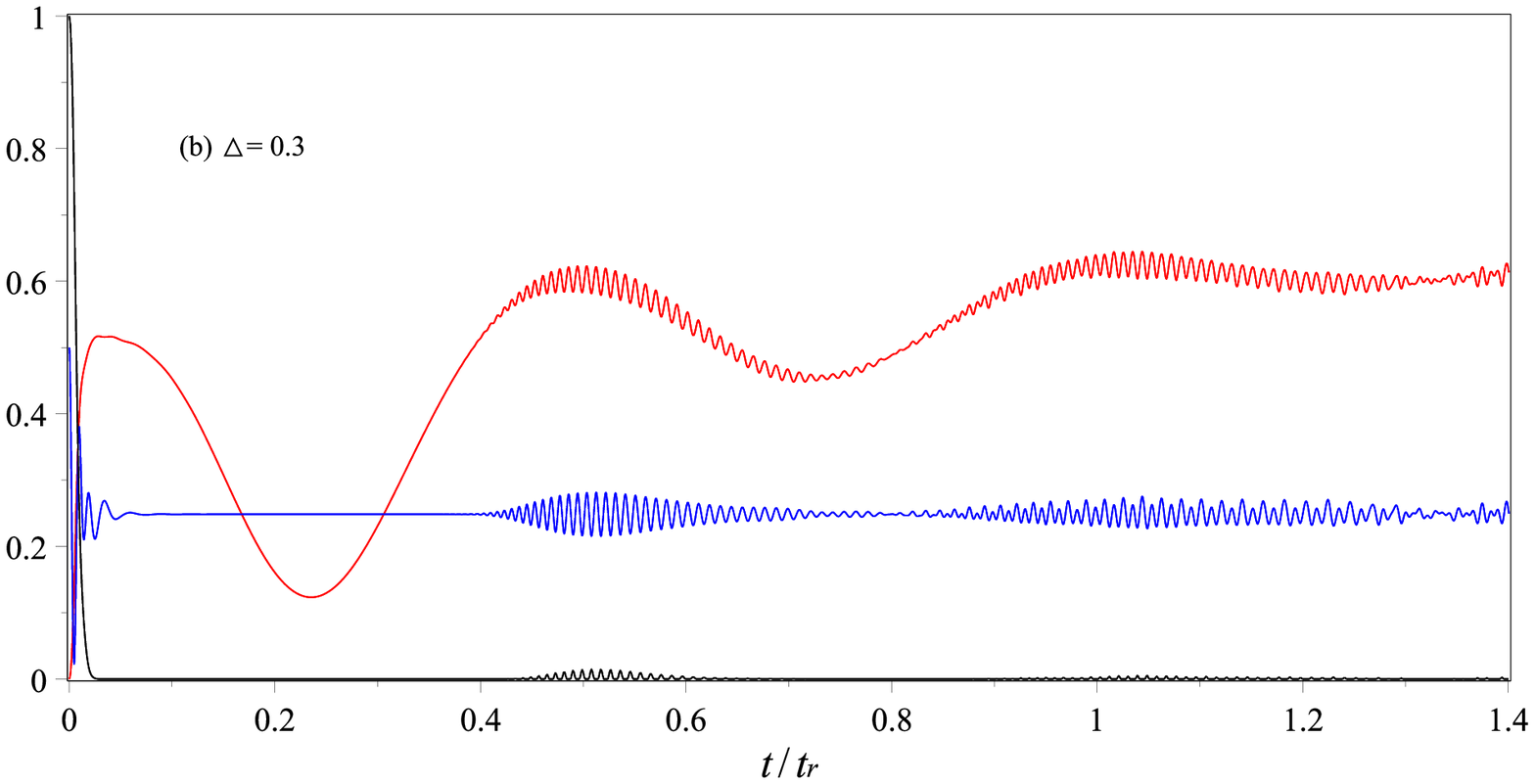}%
		}
		
	\subfloat[$\Delta=0.5$]{%
		\includegraphics[width=0.85\columnwidth]{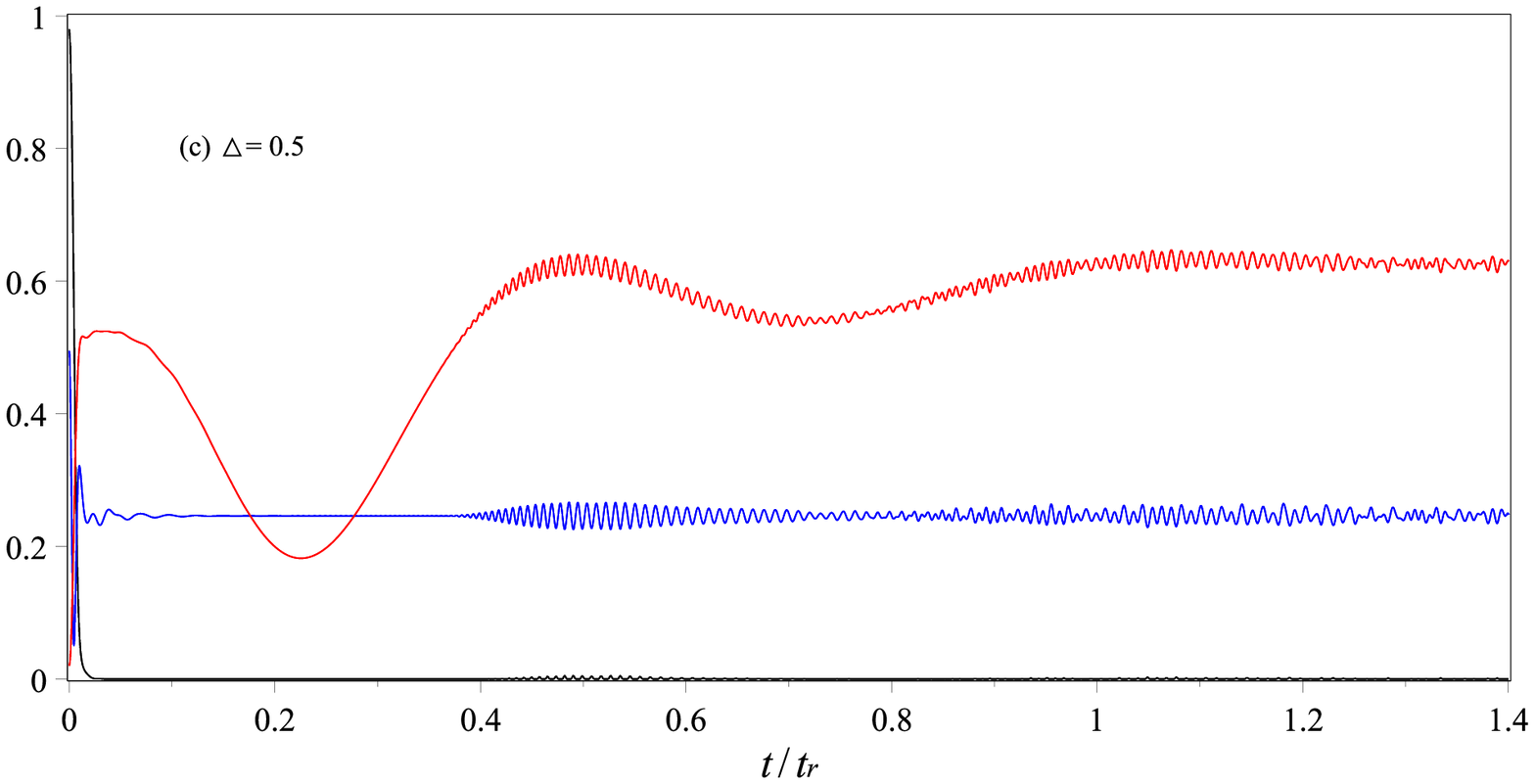}%
		}
	
\caption{Plots comparing qubits linear entropies (red), probability of the two qubits state $\left|ee\right\rangle$ (blue) and tangle (black) for two qubits-composite spin models of initial qubit state $\frac{1}{\sqrt{2}}(\left|ee\right\rangle+\left|gg\right\rangle)$, $|\zeta|^2=25$, $N=150$ and the composite spin initial phase of $\phi=0$, with decoherence effects.  Figure (a) shows the differences in the system with $\Delta = 0.1$, (b) shows the differences in the system with $\Delta = 0.3$ and (c) shows the
differences in the system with $\Delta = 0.5$.}
\label{fig:DCGBS}
\end{figure}

\begin{figure}[h]
	\subfloat[$\Delta=0.1$]{%
		\includegraphics[width=0.85\columnwidth]{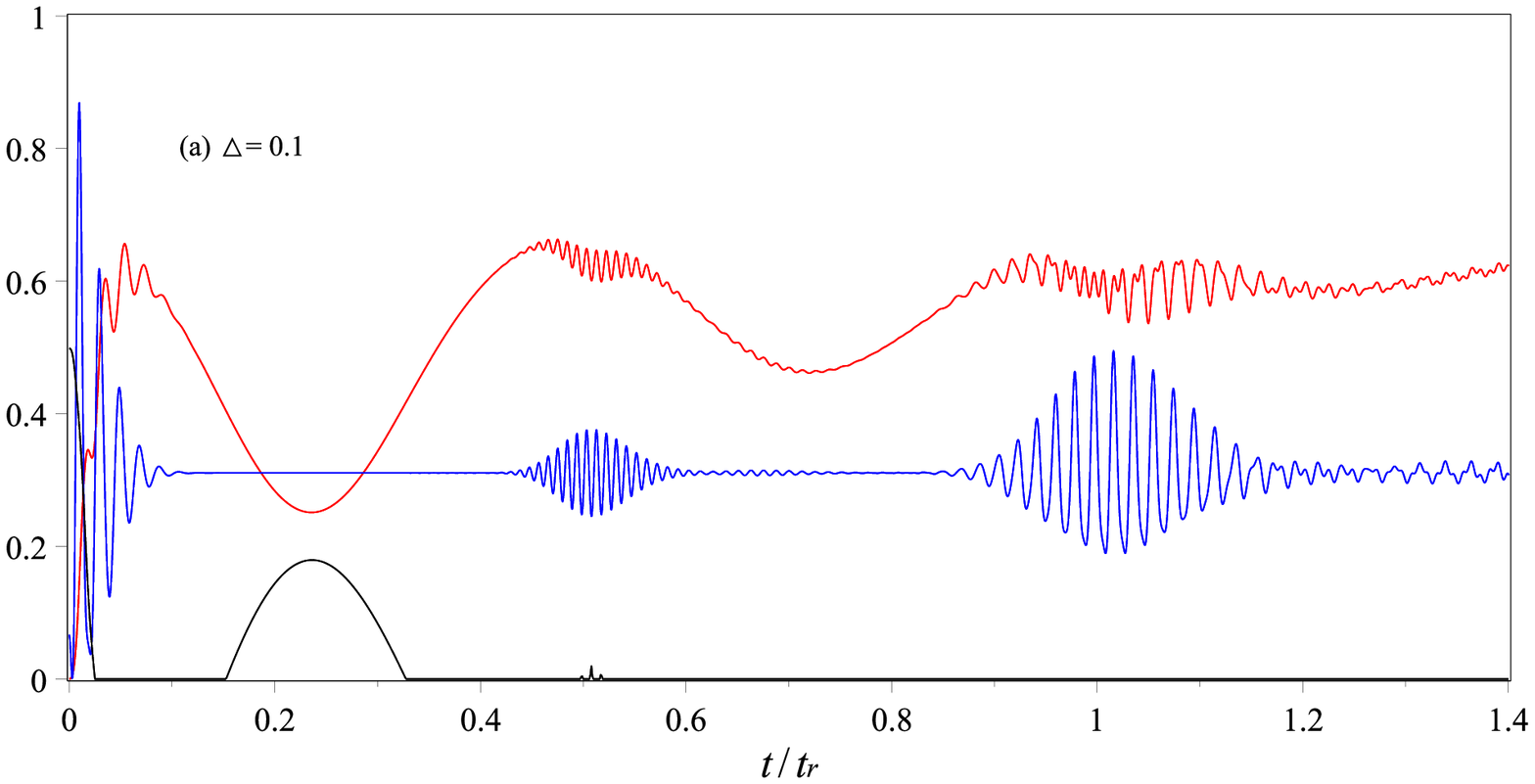} %
		}
		
	\subfloat[$\Delta=0.3$]{%
		\includegraphics[width=0.85\columnwidth]{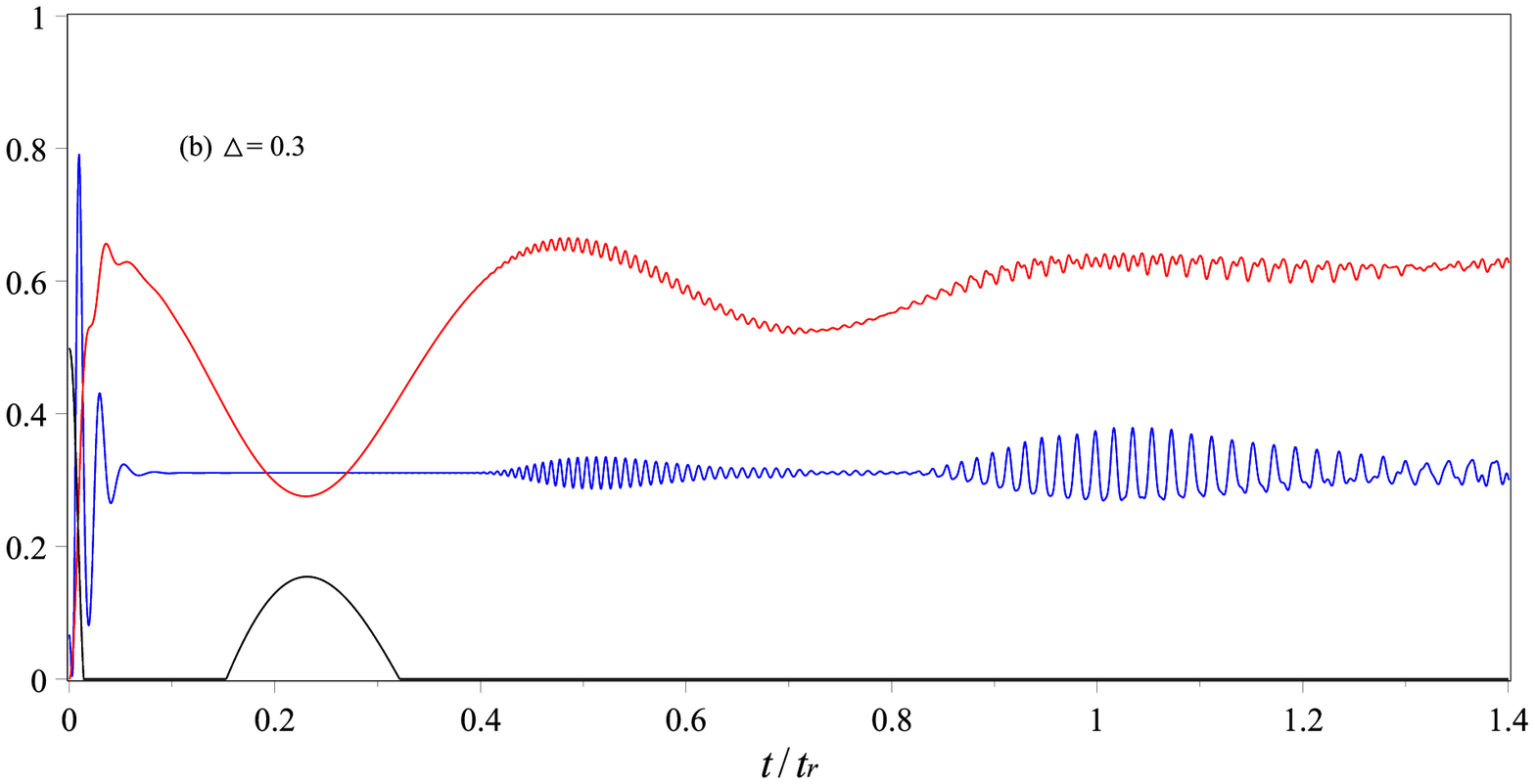}%
		}
		
	\subfloat[$\Delta=0.5$]{%
		\includegraphics[width=0.85\columnwidth]{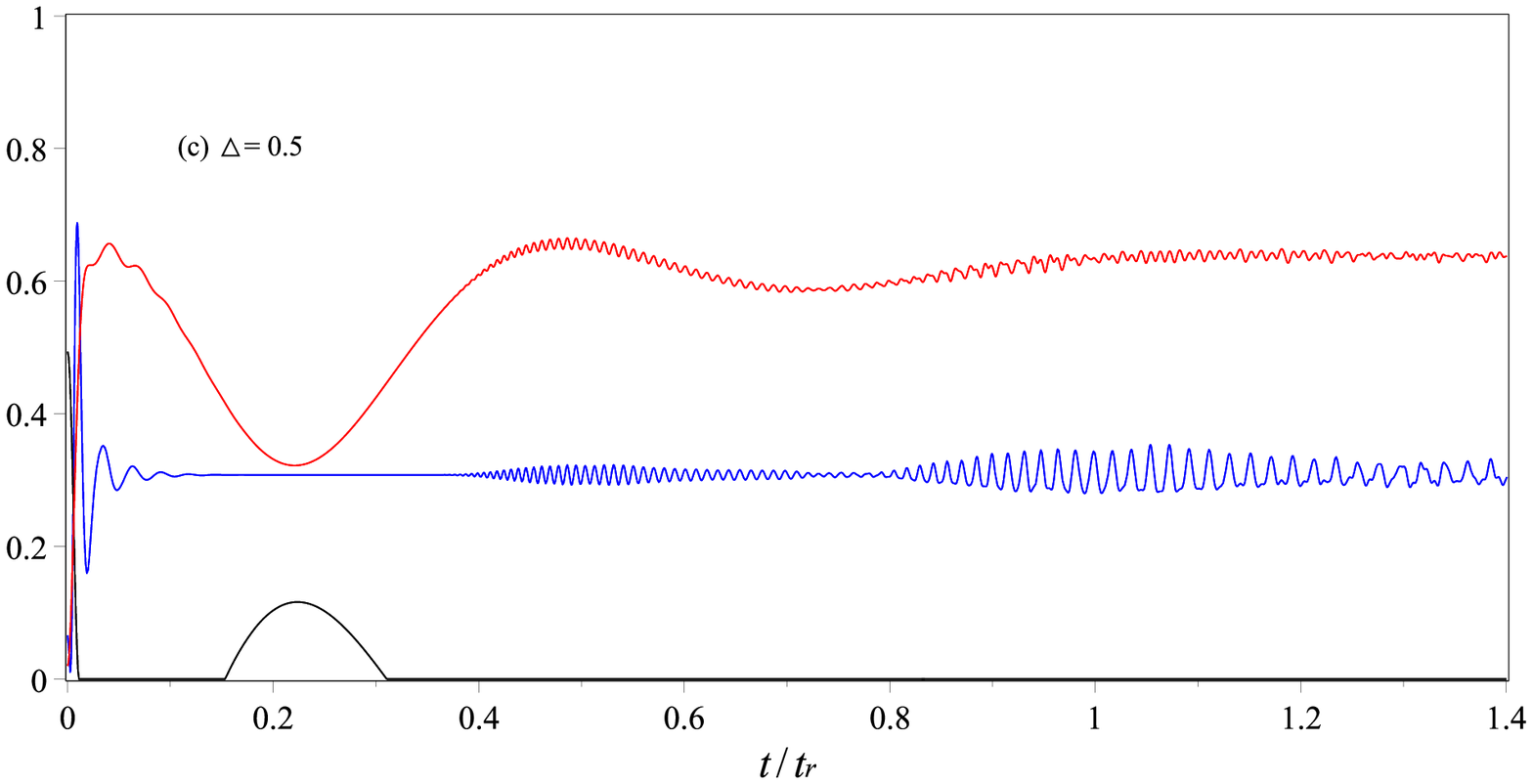}%
		}
	
\caption{Plots comparing qubits linear entropies (red), probability of the two qubits state $\left|ee\right\rangle$ (blue) and concurrence (black) for two qubits-composite spin models of initial qubit state $\frac{1}{\sqrt{15}}\left|ee\right\rangle+\sqrt{\frac{14}{15}}\left|gg\right\rangle$, $|\zeta|^2=25$, $N=150$ and the composite spin initial phase of $\phi=0$, with decoherence effects.  Figure (a) shows the differences in the system with $\Delta = 0.1$, (b) shows the differences in the system with $\Delta = 0.3$ and (c) shows the
differences in the system with $\Delta = 0.5$.}
\label{fig:DCGBSNM}
\end{figure}

\noindent We can also consider a discrete approximation to this ensemble of systems  which can be given in the form of

\begin{align}
\rho_q (t) \approx \sum_{\delta_i} \frac{f\left(\delta_i |0, \Delta \right) \rho_q(t,\delta_i)}{\sum_{\delta_i}f\left(\delta_i |0, \Delta \right)}
\end{align}

\noindent where $i$ indicates the number of the discrete events and $\Delta $ is still the distribution width. As this number $i$ increases, the function tends towards resembling a normal distribution in the continuous regime. When using the discrete approximation to numerically evaluate approximations to density matrices, values of $\delta$ are sampled over a range  from $-3\Delta$ to $3\Delta$, with $i$ of sufficient size to provide accurate approximations.
Fig.~\ref{fig:DCGCS} and Fig.~\ref{fig:DCGBS} respectively visualize the comparisons between relevant features in a two qubits-field mode and two qubits-composite spin model, with decoherence effects of varying the distribution width $\Delta$ value, for the case of the initial two-qubit state inside the `basin of attraction'. Fig.~\ref{fig:DCGBSNM} also depicts similar features for a two qubits-composite spin model with decoherence effects, but considering an initial two-qubit state that lies outside the `basin of attraction'. Significant differences can be observed in the qubits' probability, linear entropy and entanglement values as the amount of decoherence increases, whilst both systems exhibit similar attributes. One notable change  can be seen through the decreasing amplitude of the qubits' probability at the revival time and its half. It is also very interesting to note that at the attractor times, the values of linear entropy have increased which suggests that there is an effect of decoherence on the purity of the system; however, the qubits remain unentangled as $\tau$ remains zero. Decoherence however shows its effects on the initial decay of the measurement, where entanglement disappears more rapidly and revives with decreased amplitudes as the errors increase.

\section{Two Qubits-Composite Spin System with Small $N$}

\begin{figure}[h]
	\subfloat[no decoherence effects]{%
		\includegraphics[width=0.85\columnwidth]{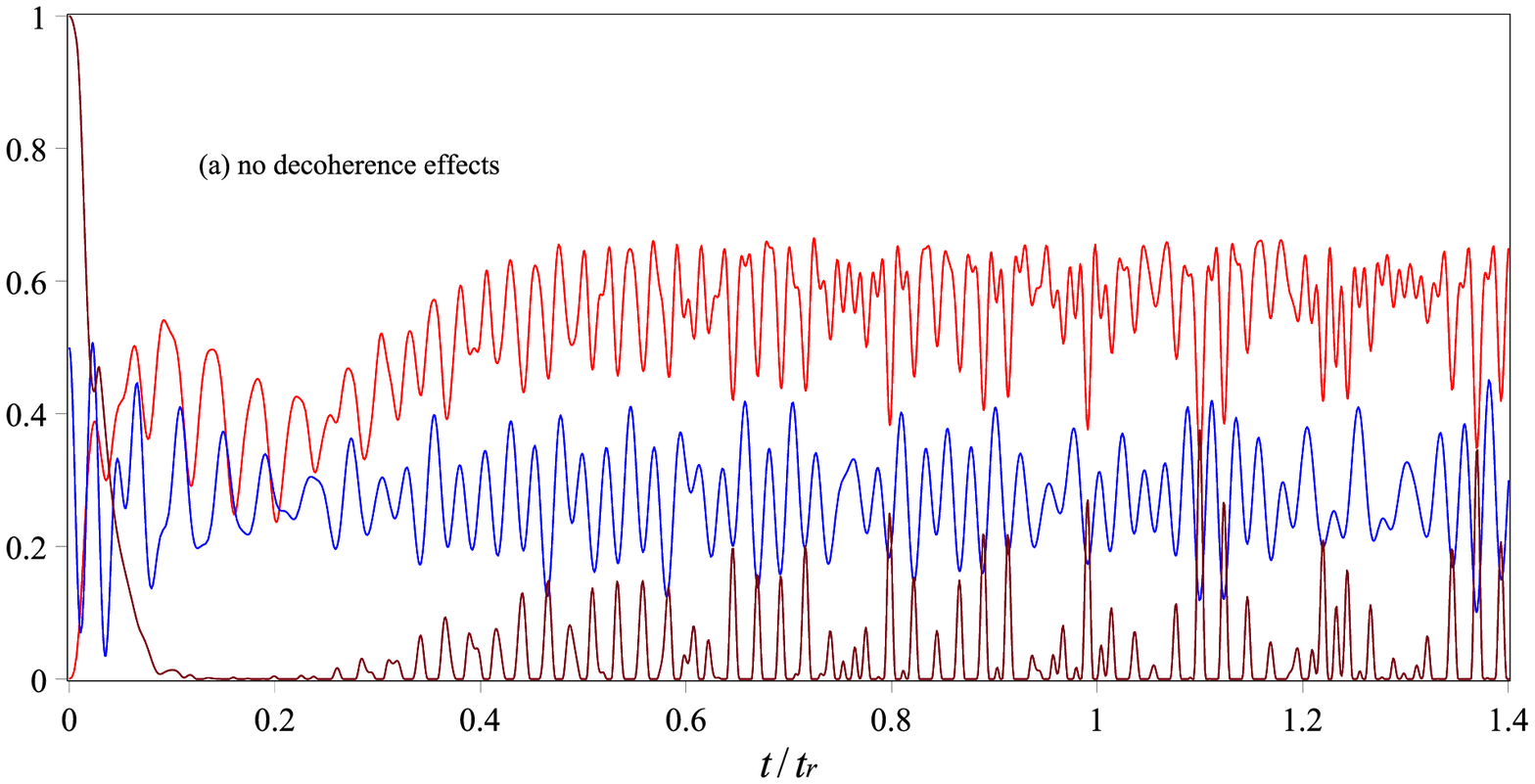} %
		}
		
	\subfloat[$\Delta=0.1$]{%
		\includegraphics[width=0.85\columnwidth]{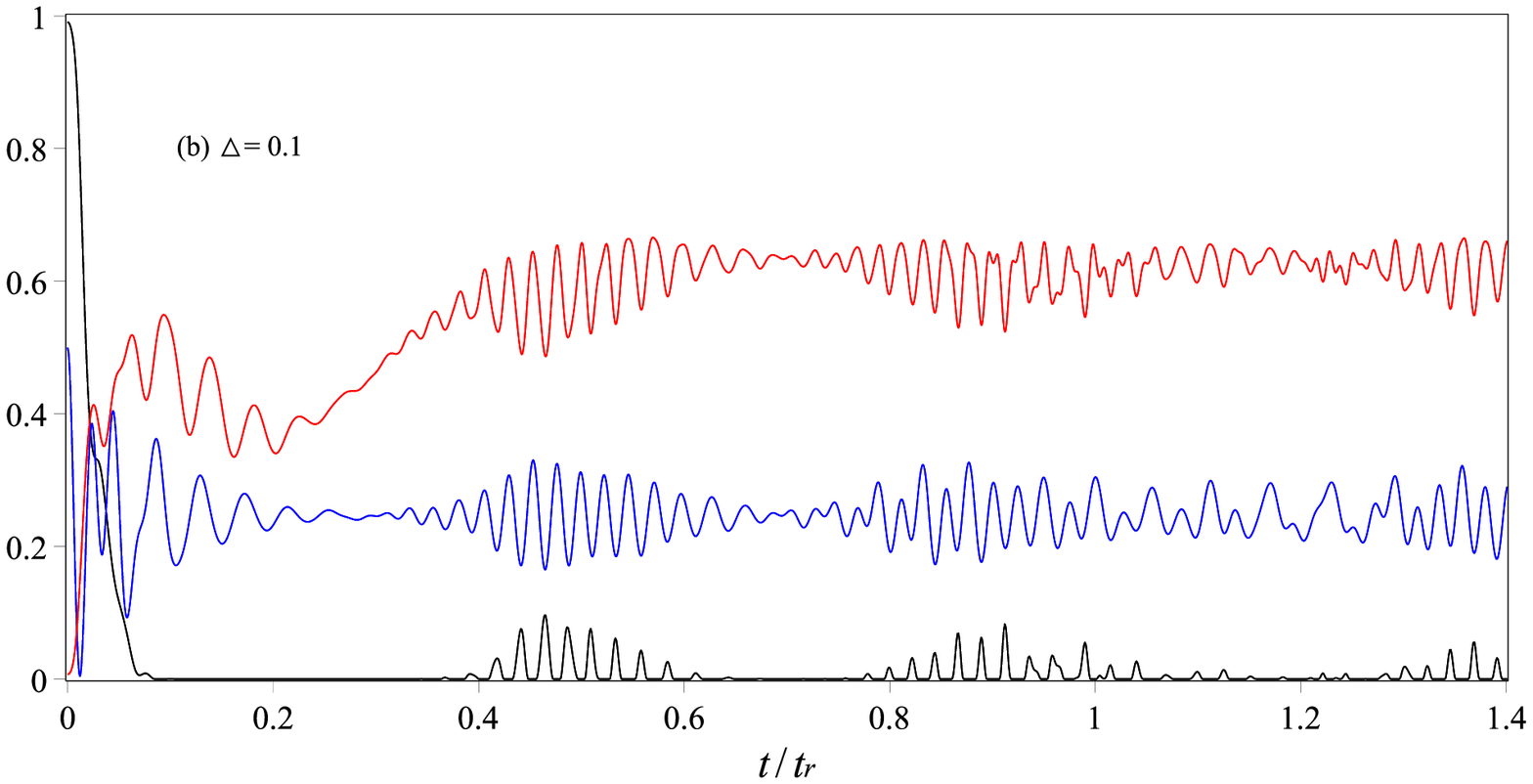}%
		}
		
	\subfloat[$\Delta=0.3$]{%
		\includegraphics[width=0.85\columnwidth]{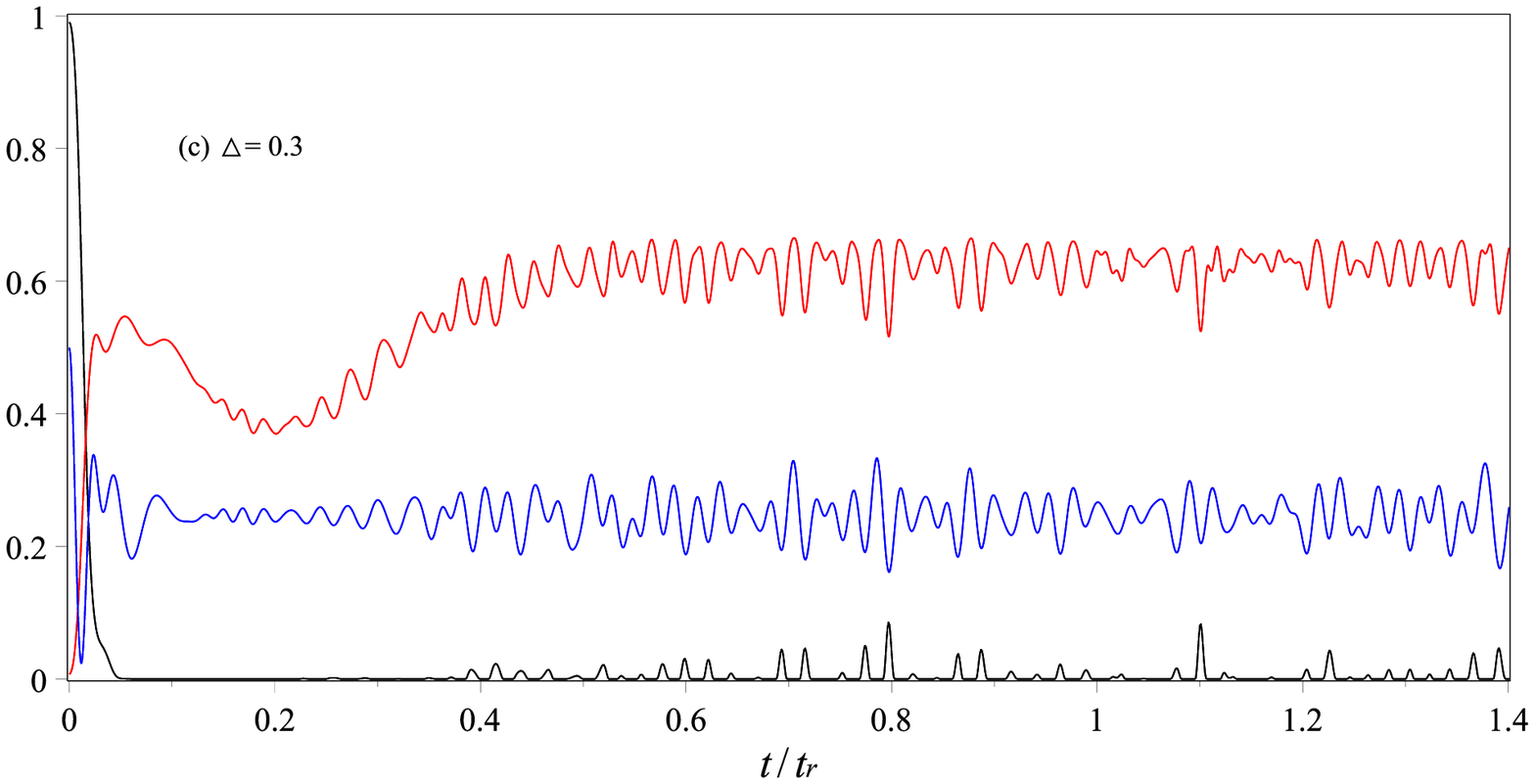}%
		}
	
\caption{Plots comparing qubits linear entropies (red), probability of the two qubits state $\left|ee\right\rangle$ (blue) and tangle (black) for two qubits-composite spin models of initial qubit state $\frac{1}{\sqrt{2}}(\left|ee\right\rangle+\left|gg\right\rangle)$, $|\zeta|^2=9$, $N=25$ and the composite spin initial phase of $\phi=0$, with decoherence effects.  Figure (a) shows the system with no decoherence effects, (b) shows the differences in the system with $\Delta = 0.1$ and (c) shows the differences in the system with $\Delta = 0.3$.}
\label{fig:DCGBSN}
\end{figure}

The calculations we have presented thus far have been performed with $N=150$ and a spin coherent state parameter $|\zeta|^2=25$, conditions such that $N$ is large enough to make a good comparison with the field mode case and to enable comparison with previously published work on this model. However, in order to consider the potential for experiments that could be realistic with current or near future systems, it is interesting to consider smaller values for $N$.
We have therefore extended our study by exploring the regime of small qubit number in the composite spin. We investigated to ascertain the smallest $N$ that maintains the occurrence of collapse and revival of entanglement in the two-qubit system. Some signature for this phenomenon can be seen for $N \sim 20$ and this is more apparent in the case $N$ equal to 25 and $|\zeta|^2=9$, as presented here. The time evolutions of this system with the effects of different value of decoherence are plotted as Figure \ref{fig:DCGBSN}.

From this figure, we can see that the collapse and revival of the Rabi oscillations, as well as the linear entropy of the system, are rather noisy in comparison to the larger $N$ results. However, there is still clear evidence for revival of the entanglement between the two qubits. It is also very interesting to note that even with a modest amount of decoherence at $\Delta=0.1$, the collapse and revival of entanglement in the qubits system can still be observed. The entanglement does, however, collapse completely with an error distribution width larger than $\Delta=0.5$.  

\section {Conclusions}

\noindent We have demonstrated that collapse and revival of the entanglement between two qubits interacting with a composite spin system exhibits similar features to the previous demonstrations of these phenomena in two-qubit systems coupled to a field mode, when the composite spin contains a sizeable number ($N=150$) individual spins. Analogous to the field mode scenario, in the composite spin scenario there is also a `basin of attraction' for initial two-qubit states to show this collapse and revival, and for states outside this basin the related phenomenon of sudden death and birth is seen. Our studies are relevant for physical implementations where two specific qubits couple to a set of other qubits, as could be realised in Circuit QED \cite{mcdermott2005simultaneous, niskanen2007quantum, neeley2010generation, neeley2009emulation, dicarlo2010preparation, reed2012realization,kirchmair2013observation}, or ion/atom \cite{bohnet2016quantum, reiter2017autonomous, schindler2013quantum, gorman2017quantum}, or molecular \cite{jones2009magnetic,simmons2010magnetic}, or quantum dot \cite{reilly2010exchange, chekhovich2013nuclear} systems.

In any physical realisation of a system of two qubits coupled to a field mode, or to a composite spin, there will clearly be the potential for errors and thus decoherence. One important example that we have studied here is the case of a mismatch in the (supposedly identical) couplings of the two qubits, which might for example arise due to spatial positioning errors for physically identical fundamental qubits, or to fabrication and positioning errors in the case of manufactured qubits. We have considered coupling errors subject to a distribution, leading to additional mixture in the two-qubit reduced density matrix, for both the case of a field mode and a composite spin coupled to the two-qubit system. In both of these scenarios the decoherence due to coupling errors reduces the strength of entanglement revivals. However, the promising observation is that for a 10 \% error ($\Delta =0.1$) in coupling there are still very significant entanglement revival or rebirth signatures. This is encouraging from the perspective of experimental investigations of such phenomena, where errors will be present. To further encourage experimental investigations, we have also explored the composite spin parameter space at much smaller values of $N$, which are likely to be more readily accessible with current or near future implementations. We have illustrated this with the example of $N=25$, where there is clear evidence for entanglement revival and robustness against 10 \% coupling errors.

Clearly there are other forms of error and thus decoherence that can be considered. As an example, future work will address a mismatch in frequency of the two qubits, which could arise due to identical qubits experiencing slightly different external conditions, or to fabrication errors in manufactured qubits.
 
\addcontentsline{toc}{chapter}{\sc Bibliography}
\bibliography{qubitbigspin}
\bibliographystyle{ieeetr}
\end{document}